\documentclass[aps,amsmath,amssymb,prb,twocolumn,showpacs]{revtex4}

\usepackage{amsfonts}
\usepackage{amsmath}
\usepackage{amssymb}
\usepackage{graphicx}
\usepackage{color}

\newcommand{\rvec}{\mathbf{r}}

\newcommand{\kvec}{\mathbf{k}}
\newcommand{\avec}[1]{\mathbf{#1}}

\newcommand{\Del}{\nabla}
\newcommand{\del}{\nabla}
\newcommand{\beq}{\begin{equation}}
\newcommand{\eeq}{\end{equation}}
\newcommand{\bsp}{\begin{split}}
\newcommand{\esp}{\end{split}}
\newcommand{\bpm}{\begin{pmatrix}}
\newcommand{\epm}{\end{pmatrix}}

\newcommand{\eq}[1]{Eq. (\ref{#1})}
\newcommand{\fig}[1]{Fig. \ref{#1}}
\newcommand{\tab}[1]{Table \ref{#1}}

\begin{document}

\title{Quantum oscillations and Berry's phase in topological insulator surface states with broken particle-hole symmetry}

\author{Anthony R. Wright}
\email{a.wright7@uq.edu.au}
\affiliation{School of Mathematics and Physics, University of Queensland, Brisbane, 4072 Queensland, Australia}
\author{Ross H. McKenzie}
\affiliation{School of Mathematics and Physics, University of Queensland, Brisbane, 4072 Queensland, Australia}

\date{\today}

\begin{abstract}
Quantum oscillations can be used to determine
properties of the Fermi surface of metals by varying the magnitude and orientation of an external magnetic field. Topological insulator surface states are an unusual mix of normal and Dirac fermions. Unlike in graphene and simple metals,  Berry's geometric phase in topological insulator surface states is not necessarily quantized. 
We show that reliably extracting this geometric phase 
from the phase offset associated with the quantum oscillations
is subtle.
This is especially so in the presence of a Dirac
gap such as that associated with the Zeeman splitting or
interlayer tunneling.
We develop a semi-classical theory for general mixed normal-Dirac systems in the presence of a gap, and in doing so clarify the role of topology
and broken particle-hole symmetry.
We propose a systematic procedure of fitting Landau level index plots at large filling factors
to reliably extract the phase offset associated with Berry's phase.
\end{abstract}


\maketitle

\section{Introduction}

Topological insulators are bulk insulators with  metallic surface states that are topologically protected \cite{kane, kanerev, zhangrev}. 
The quantized Hall resistance in two-dimensional topological insulators was measured several years ago \cite{koenig, roth1}.
Recently measurements have been made of properties of the surface states of three-dimensional topological insulators.
Angle Resolved PhotoEmission Spectroscopy (ARPES) studies
provide direct imaging of near--surface bands \cite{hussain, hasan1}
showing dispersions of  an odd number of Dirac cones,
as expected for a topological insulator. 
Furthermore, these band structures are consistent with the presence of a non-trivial Berry's phase of $\pi$.\cite{hasan0}  
However, these surface band structures also exhibit significant band-bending, and particle-hole symmetry with respect to
the Dirac point is broken.

A complementary method to ARPES to determine the Fermi surface properties of a 
metal is to measure quantum oscillations such as associated with
the Shubnikov de Haas or de Haas van Alphen effects.\cite{CM} 
Such experiments  can distinguish between bulk and surface states, even if both are gapless (or near-gapless), by tilting the magnetic field  \cite{qo1}. 
Furthermore, the phase offset ($\gamma$) of the quantum oscillations 
is related to the Berry's phase associated with cyclotron orbits
and provides a means to experimentally access this important signature of a Dirac cone. Experiments on graphene \cite{novogeim, Zhang}
found evidence of the expected non-trivial Berry's phase of $\pi$.
There have now been a number of quantum oscillation experiments on 3D topological insulators.\cite{qo1, qo2, natphys, ren, ando3, mol, veldhorst, nanoL,xiong, xiong2, natcom,ando33,ren2} A prevalent observation in these studies is that the phase offset is not equal to the zero value associated with a Berry's phase
of $\pi$. In Table \ref{gammatable} we have listed the different phase offsets obtained by various groups and materials. The deviation of the observed $\gamma$ from zero has been variously attributed to the Zeeman effect \cite{natphys, SWP}, and the non-ideal Dirac (i.e. nonlinear) nature of the surface states.\cite{ando} In the present work, we reformulate both of these proposals in a precise and concrete manner.

\begin{table}[h]
\centering
\begin{tabular}{c | c | c | c}
  \hline  \hline                      
Material & $B_0 (T)$ & $\gamma$ & Ref. \\ \hline
Bi$_{0.91}$Sb$_{0.01}$ & 0.65 & 0.25 & \onlinecite{qo2}\\
Bi$_{2}$Se$_{3}$ (S2) & 32 & -0.5 & \onlinecite{natphys}\\
Bi$_{2}$Se$_{3}$ (S3) & 100 & -0.7 & \onlinecite{natphys}\\
Bi$_{2}$Te$_{2}$Se & 64 & $0.22\pm0.12$ & \onlinecite{ren}\\
Bi$_{2}$Te$_{3}$ & 50 & $-0.05\pm0.05$ & \onlinecite{veldhorst}\\
Bi$_{2}$Se$_{3}$ & 16 & $-0.15\pm0.08$ & \onlinecite{nanoL}\\
Bi$_{2}$Te$_{2}$Se (S1) & 60 & $0.05\pm0.02$ & \onlinecite{xiong}\\
Bi$_{2}$Te$_{2}$Se (S4) & 47 & 0.32 & \onlinecite{xiong}\\
Bi$_{2}$Te$_2$Se & 73 & -0.05 & \onlinecite{xiong2}\\
Bi$_{2}$Se$_3$ & many & $-0.1\pm0.1$ & \onlinecite{natcom}\\
Bi$_{1.996}$Sn$_{0.004}$Te$_2$Se & 116 & $-0.1\pm0.1$ & \onlinecite{ren2}\\
graphene & 44 & 0 & \onlinecite{novogeim}\\
  \hline  \hline
\end{tabular}
\caption{The different phase offsets ($\gamma$) measured for various materials and at various chemical potentials (quantified by the frequency of the quantum oscillations ($B_0$), the surface area of the cyclotron orbit in teslas). The second last entry was performed at more than 20 different chemical potentials. S$\#$ denotes the sample number in the experiment.}
\label{gammatable}
\end{table}

The aim of this Article is to provide a quantitative framework 
for using quantum oscillation experiments to characterise topological insulator surface states. 
In particular, we focus on extracting the
phase offset and the cyclotron effective  mass from
experiments. 
For mixed normal-Dirac systems (i.e. Dirac cones with a particle-hole symmetry breaking quadratic term), we find that there is a key energy scale relevant to quantum oscillation experiments: the normal fermion mass multiplied by the Dirac Fermi velocity squared, $mv_F^2$. 
The interplay of this energy scale with the Zeeman or intrinsic Dirac-gap ($\Delta$)
 introduces subtleties to quantum oscillations that both complicate and enrich the observed phenomena. In particular, we find that the phase offset is only quantized
if the system is either gapless or particle-hole symmetric. 
In a topological insulator surface state under an applied magnetic field, the phase offset is \emph{never} independent of magnetic
field  because of the Dirac mass gap associated with the Zeeman splitting, together with the ubiquitous particle-hole symmetry breaking. 
We outline how one can circumvent this difficulty: by fitting the experimental data at large filling factors to a simple function (\eq{largen}), one can obtain a linearized form which is asymptotically exact in zero field. The intercept of such a linear plot as $1/B\rightarrow 0$ yields the topologically relevant phase offset at zero field. 
We hope that this analysis will be valuable to future experimental studies, and will stimulate further studies utilising quantum oscillation experiments to investigate
other topological insulator regimes, such as associated with
thin films.\cite{moore1, moore2}

The structure of the article is as follows. In Sec. II, we develop the semiclassical theory of quantum oscillations when particle-hole symmetry is broken. In Sec III, we introduce a specific model
Hamiltonian. 
In Sec. IV, we analyse the quantum oscillations in the model system, and focus on the non-universal phase offsets and the subtleties of cyclotron effective mass measurements. This section ends with an analysis of expected Landau level index plots in these experiments. In Sec. V, we propose a robust procedure to extract the phase offset from experiment, overcoming the difficulties which led to the wide range of values in \tab{gammatable}. In Sec. VI we compare our results with existing experiments and show that the procedure outlined in the preceding section can substantially improve phase offset measurements.  In Sec. VII, we comment briefly on the relevance of our results to spintronic systems.


\section{Semiclassical theory of quantum oscillations}
By measuring the longitudinal resistivity of a metal sample as a function of external field strength $B$, one observes that it oscillates according to\cite{CM}

\beq
\Delta\rho_{xx} \propto \cos\biggl[2\pi\biggl(\frac{B_0}{B} - \gamma\biggr)\biggr],
\label{resistivity}
\eeq
where the oscillation frequency, $B_0$, is related to the area of the
Fermi surface enclosed by the cyclotron orbit, and $\gamma$ is a phase offset. This is the Shubnikov--de-Haas effect. The de-Haas--van-Alphen effect, which is the oscillation of the magnetisation of a sample as a function of $B$ takes a similar form, and in two dimensions is given by \eq{M}. The phase offset, $\gamma$ is related to the Berry's phase of the cyclotron orbit.\cite{Mx} A key issue we clarify in the present work is the role of the Berry's phase in the phase offset $\gamma$.

\subsection{The phase offset in quantum oscillations}
Under a weak external magnetic field, electrons follow equi-energy contours about the electronic dispersion. In order to construct a semi-classical theory of this situation, one usually begins with Onsager's semi-classical quantisation condition due to the single-valuedness of Bloch wavefunctions \cite{roth}. The k-space area $S(C)$ of a closed electronic
orbit $C$ is quantized
according to 
\beq
S(C)l_B^2 = 2\pi(n + \gamma(C))
\eeq
where 
$l_B$ is the magnetic length, defined by $l_B = \sqrt{\hbar/eB}$,
and $n$ is an integer.
The phase offset $\gamma(C) = 1/2 - \Gamma(C)/2\pi$, with $\Gamma(C)$ the Berry's phase of the orbit $C$ \cite{berry, Niu}, given by
\beq
\Gamma_\alpha(C) = 
i\oint_Cd\kvec\cdot \langle u_{\kvec,\alpha}|\del_{\kvec}u_{\kvec,\alpha}\rangle.
\label{berry}
\eeq
in the band $\alpha$. This  is expressed in terms of the eigenvectors $|u_{\kvec,\alpha}\rangle$ of the $k-$dependent Hamiltonian

\beq
H(\kvec) = \exp(-i\kvec\cdot\rvec)H \exp(i\kvec\cdot\rvec),
\eeq
where $H$ is, say, a tight-binding Hamiltonian. For normal fermions with a quadratic dispersion, or in fact any isolated band, $\Gamma(C) = 0$ for any contour $C$. For massless Dirac fermions, the Berry's phase is again path independent, giving $\Gamma(C) = \pi$. Accounting only for the bare dispersion contribution to the cyclotron orbit, it is tempting to take $C$ to be the equi-energy closed path of the zero-field dispersion. Thus $\gamma(C)$ for a contour which includes only the bare dispersion contribution to the energy gives a constant $\gamma = 1/2$ for normal fermions, and $\gamma = 0$ for massless Dirac fermions \cite{MS1999}. However, for massive Dirac fermions the dispersion is no longer linear at low energies, and the Berry's phase is no longer path independent, (i.e. $\Gamma = \Gamma(C)$), and so $\gamma(C)$ is not, in general, quantized.\cite{Mx}

In a clear and insightful recent article, Fuchs \emph{et al.} \cite{Mx} pointed out that for a massive Dirac cone in a magnetic field, there is a pseudo-spin orbital magnetic moment contribution to the energy that must be taken into account for cyclotron orbits (\emph{i.e.} equi-energy contours). The Onsager condition for cyclotron orbits is generalised in this case to

\beq
S(\epsilon)l_B^2 = 2\pi(n + \gamma_\alpha(\epsilon)).
\label{S}
\eeq
The quantity $\gamma_\alpha(\epsilon)$ is not, in general, equal to $\gamma(C)$, since it also includes the pseudo-spin magnetisation component. It is $\gamma_\alpha(\epsilon)$ that gives the relevant phase offset, and not $\gamma(C)$ in \eq{S}.

Somewhat surprisingly, for the particle-hole symmetric massive Dirac cone, this extra contribution to the phase offset due to the pseudo-spin orbital magnetic moment exactly cancels the energy-dependent part of the Berry's phase, such that $\gamma_\alpha(\epsilon) = 0$ after all.\cite{Mx} By inverting \eq{S}, it can be shown that the semi-classical Landau levels exactly reproduce the fully quantum mechanical ones in the specific case of a massive Dirac cone. For this reason, Fuchs \emph{et al.} label $\gamma_\alpha(\epsilon)$ ($\gamma_L$ in their nomenclature) the `Landau index shift', and they associate it with a winding number of the orbit -- a topological quantity. Therefore, the phase offset  is strictly $\gamma = 1/2$ for normal fermions, and $\gamma = 0$ for particle-hole symmetric Dirac fermions.

\subsection{A general expression for the phase offset in quantum oscillations: broken particle-hole symmetry} 
A further goal of this work is to generalise the contribution of Fuchs \emph{et al.} to systems with broken particle-hole symmetry. This extension applies to many systems, most notably at present, topological insulator surface states \cite{Bi1,Bi2}. We find that in the case of broken particle-hole symmetry, the winding number can no longer be associated with the magnetic moment-adjusted Berry's phase, and so the strict quantisation of $\gamma_\alpha(\epsilon)$ is removed.

Consider a general two band system. Following Fuchs \emph{et al.},\cite{Mx} the energy of an electron in band $\alpha$ is 

\beq
\epsilon_\alpha(\kvec) = \epsilon_{0,\alpha}(\kvec) - \mathcal{M}_\alpha(\kvec)\cdot\mathcal{B},
\eeq
where $\epsilon_{0,\alpha}(\kvec)$ is the bare dispersion energy in zero field, $\mathcal{B}$ is the external magnetic field, and the pseudo-spin orbital magnetic moment \cite{Niu,Mx}

\beq
\bsp
\mathcal{M}_\alpha(\kvec) &=\frac{e}{2}\langle(\avec{\hat{r}}-\avec{r_c})\times\avec{\hat{j}}\rangle\\
&= \frac{e}{2\hbar}(\epsilon_{0,\alpha}(\kvec) - \epsilon_{0,\bar\alpha}(\kvec))\Omega_\alpha(\kvec),
\end{split}
\label{MM}
\eeq
where $\avec{r}_c$ is the centre of mass position of the wavepacket, $\avec{\hat{j}}$ is the current operator, and $\bar\alpha$ denotes the band which is not $\alpha$. We emphasise that \eq{MM} is a property of Bloch electrons in a magnetic field, and has nothing to do with spin. $\Omega_\alpha(\kvec)$ is the Berry curvature which is the local quantity given by

\beq
\Omega_\alpha(\kvec) = \avec{\Del}_\kvec\times i\langle u_{\kvec,\alpha}|\avec{\Del}_\kvec u_{\kvec,\alpha}\rangle
\eeq
and is related to the Berry's phase by Stokes' theorem such that
\beq
\Gamma_\alpha(C) = \int_S \Omega_\alpha(\avec{k})d\avec{k},
\label{Gam2}
\eeq
where $S$ is the area enclosed by $C$. \eq{MM} generalizes Appendix C in Ref.(\onlinecite{Mx}) to the case of broken particle-hole symmetry. Including the magnetic moment in \eq{S}, we can obtain the full quantisation condition for a generic two-band model, by finding

\beq
\bsp
S(\epsilon_{0,\alpha}&(\kvec))l_B^2 = S(\epsilon_{\alpha}(\kvec) + \mathcal{M}_{\alpha}(\kvec)\cdot\mathcal{B})l_B^2\\
& \approx S(\epsilon_{\alpha}(\kvec))l_B^2 + \frac{1}{2}(\epsilon_{\alpha}(\kvec) - \epsilon_{\bar\alpha}(\kvec))\frac{d\Gamma_{\alpha}(\epsilon)}{d\epsilon_{\alpha}(\kvec)}.\\
\end{split}
\eeq
In the last equality we made use of \eq{MM} and \eq{Gam2}.
Rearranging the above expression to obtain $S(\epsilon_\alpha(\kvec))$, and recognising that we already know $S(\epsilon_{0,\alpha}(\kvec))$ from Onsager's condition, \eq{S} we obtain
\beq
\gamma_\alpha(\epsilon) = \frac{1}{2} - \frac{1}{2\pi}\biggl[\Gamma_\alpha(\epsilon) +  \frac{1}{2}(\epsilon_{0,\alpha} - \epsilon_{0,\bar\alpha})\frac{d\Gamma_\alpha(\epsilon)}{d\epsilon_{0,\alpha}}\biggr].
\label{gam}
\eeq
This result for the phase offset in particle hole symmetry broken mixed Dirac-normal fermion systems directly leads to \eq{gam2}, the central result of this paper.
Two limiting cases are worth mentioning. Firstly, for particle-hole symmetric Hamiltonians, $\epsilon_{0, \bar\alpha} = -\epsilon_{0, \alpha}$, and the second and third terms on the right together reduce to a multiple of the winding number already obtained by Fuchs \emph{et al.}, which is simply $1/2$, giving $\gamma_\alpha(\epsilon) = 0$. Secondly, for a degenerate two band system, $\epsilon_{0,\bar\alpha} = \epsilon_{0,\alpha}$, and $\Gamma = 0$, trivially giving $\gamma_\alpha(\epsilon) = 1/2$. In general, however, no such simplification is possible.

\subsection{Semi-classical theory of the de-Haas--van-Alphen effect}
The Lifshitz-Kosevich theory of the de-Haas van-Alphen effect for a single species of free electrons has been obtained by Champel and Mineev \cite{CM}. Starting from an expression for the Green's function of an electron in a magnetic field, they obtain the density of states, and then the thermodynamic potential. From here one can directly obtain the magnetization, and see that it oscillates as a function of magnetic field. In particular, this demonstrates the direct relation between oscillations in density of states as a function of field strength, and how they directly lead to oscillations in magnetization. 

By inverting Onsager's condition, \eq{S}, in terms of energy levels, and thus replacing the energy levels in a magnetic field with an expression for the quantized orbits $S(\epsilon)$ and the phase offset $\gamma$, Luk'yanchuk and Kopelevich \cite{LK} generalised the treatment of the de-Haas--van-Alphen (dHvA) semiclassically, for arbitrary systems.

In a magnetic field, the electrons in band $\alpha$ undergo cyclotron orbits with frequency
\beq
\omega_c(\epsilon_\alpha) = \frac{eB}{m^*_\alpha(\epsilon)},
\eeq
where the cyclotron effective mass is
\beq
m_\alpha^*(\epsilon) = \frac{1}{2\pi}\frac{dS_{\alpha}(\epsilon)}{d\epsilon_\alpha}.
\label{mass}
\eeq

The 2D result obtained by Luk'yanchuk and Kopelevich \cite{LK} of which we are interested here, is given by
\beq
\bsp
M_{osc} &= \sum_\alpha\frac{-e}{\pi^2 \hbar^2}\frac{S_\alpha(\epsilon)}{dS_\alpha(\epsilon)/d\epsilon}\sum_{l=1}^\infty \frac{\lambda}{\sinh(\lambda l)}e^{-\frac{2 \pi l}{\omega_c \tau}}\\
& \times \sin\biggl(2\pi l\biggl[\frac{S_\alpha(\epsilon)}{2\pi \hbar e B} - \gamma_\alpha(\epsilon)\biggr]\biggr)\cos\biggl(2\pi l \frac{g_s \mu_B B}{\hbar\omega_c}\biggr)
\end{split}
\label{M}
\eeq
where 
\beq
\lambda = \frac{2 \pi^2 k_BT}{\hbar \omega_c}\frac{dS(\epsilon)}{d\epsilon}
\label{lambda}
\eeq
and $\hbar/\tau$ is the Landau level broadening.


Quantum oscillations provide an experimental method
to determine   $\gamma_\alpha(\epsilon)$. 
By assigning integers (n) to the oscillation maxima/minima in the magnetisation
or resistance as a function of inverse field,
one can extrapolate these to $1/B \rightarrow 0$ to obtain $\gamma_\alpha(\epsilon)$ which is the $n-$intercept, Modulo 1.


\section{Topological insulator surface states}
\subsection{Model Hamiltonian}
Consider a two-band Hamiltonian, where the basis is not necessarily spin, but can be thought of as a pseudo-spin, and may be spin, sub-lattice, on-site orbitals, etc, and which contains both normal fermions, and massive Dirac fermions, given by \cite{rmpspintronics, SWP, Bi1, Bi2}

\beq
H = \biggl(\frac{\hbar^2 k^2}{2m} - \mu\biggr)\mathbf{I}_2 + 
\bpm
\Delta & \hbar v_F(k_y + i k_x)\\
\hbar v_F(k_y - i k_x) & -\Delta
\epm
\label{H}
\eeq
where there are four free parameters, $m$, $\mu$, $v_F$, and $\Delta$. 
$m$ is the mass associated with the normal part of the spectrum,
$\mu$ is the chemical potential, $v_F$ is the velocity of the Dirac part of the spectrum, and $\Delta$ is the Dirac gap.

This Hamiltonian, at $\Delta = 0$, is the low energy surface theory of a time reversal invariant topological insulator. Following the classification scheme outlined by Schnyder \emph{et al.}\cite{table}, we see that the quadratic term breaks particle-hole and chiral, or sub-lattice symmetries. Since we are concerned with electrons then, time reversal is antilinear, and the subclass of the topological insulator is AII, and the topological number is a $\mathcal{Z}_2$ invariant. Generically, finite $\Delta$ gives a Dirac mass gap, and so the three dimensional system is no longer a topological insulator.

In the absence of an orbital magnetic
field there are two bands, with dispersion
\beq
\epsilon_{\alpha}(\kvec) = \frac{\hbar^2 k^2}{2m} + \alpha\sqrt{\Delta^2 + \hbar^2 v_F^2 k^2},
\label{dispn}
\eeq
where $\alpha = \pm$. 

The Landau level spectrum can be calculated exactly, giving\cite{epsn}

\beq
\epsilon_n^\alpha = \hbar \omega_0 n \pm \sqrt{2\hbar v_F^2 e B n + \biggl(\frac{\hbar \omega_0}{2} - \frac{g_s\mu_B B}{2}\biggr)^2},
\label{epsn}
\eeq
where $\omega_0 = eB/m$. In the limit $v_F\rightarrow 0$, we obtain the usual normal fermion Landau level spectrum, and in the limit $m\rightarrow \infty$, we obtain the Dirac fermion spectrum:

\begin{equation}
\epsilon_n = \left\{
\begin{array}{rl}
\hbar \omega_0(n + \frac{1}{2}) &\,\,\,\mathrm{(Normal\,\,\,fermion)}\\
\sqrt{2\hbar v_F^2 e B n} & \,\,\,\mathrm{(Dirac\,\,\,fermion)}
\end{array} 
\right.
\end{equation}
which are quoted at $\Delta = 0$. Respectively, the phase offset for quantum oscillations can be `read off', as $\gamma = 1/2$ for normal fermions, and $\gamma = 0$ for Dirac fermions. In the mixed case of \eq{epsn}, however, there is no constant which can be so readily extracted from the Landau level expression.

We can define a useful `interpolation parameter', given by 
\beq
\eta \equiv \frac{mv_F^2}{mv_F^2 + \mu},
\label{eta}
\eeq
which for $\mu,m>0$ gives $0\le \eta \le 1$. The interpolating parameter gives as two limiting cases at finite $\mu$:

\begin{equation}
\eta= \left\{
\begin{array}{rl}
0, & \quad v_F\rightarrow 0 \,\,\,\mathrm{(Normal\,\,\,fermion)} \\
1, & \quad m\rightarrow \infty \,\,\,\mathrm{(Dirac\,\,\,fermion)}.
\end{array} \right.
\label{etalims}
\end{equation}

\begin{figure}[tbp]
\centering\includegraphics[width=8.6cm]{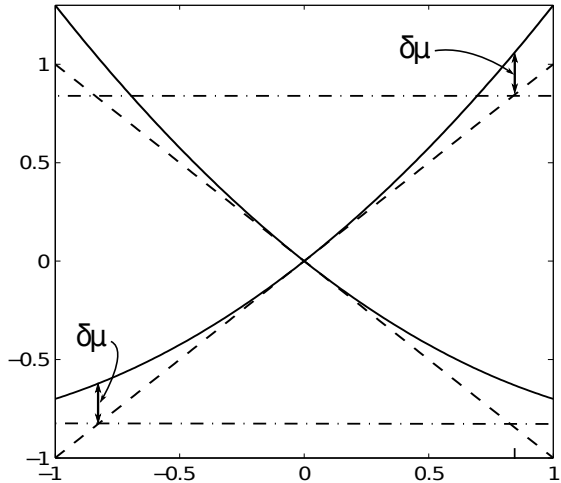}
\caption{Model band structure with particle-hole asymmetry, highlighting the effect of the $p^2/2m$ term relative to the chemical potential. In a particle-hole symmetric system, $\epsilon_+(k) = -\epsilon_-(k)$. At finite $m$, away from $k=0$, the chemical potential is shifted up for either band with respect to the symmetric case, by an amount $\delta\mu/\mu \approx \mu/(2mv_F^2) = (1/\eta - 1)/2$. In the limit $\mu\rightarrow 0$ or $m\rightarrow \infty$, the shift $\delta\mu/\mu \rightarrow 0$, restoring particle-hole symmetry.
}\label{phfig}
\end{figure}

Generally, $\eta$ small gives a Rashba spin orbit coupled 2DEG,\cite{rmpspintronics} where $v_F$ is the Rashba spin splitting, and $\Delta = g_s\mu_BB$ is the Zeeman splitting ; $\eta \approx 1$ gives, for example, graphene at $\Delta = 0$, and boron nitride at $\Delta\ne 0$,\cite{novogeim,gappedgraphene} and $0<\eta< 1$ gives the surface states of topological insulators \cite{Bi1,Bi2}, where $\Delta$ can be the Zeeman splitting or a tunnel splitting for thin films \cite{gaptheory}. Although we are most interested in the latter regime, our results are generally valid for all values of $\eta$ and $\Delta$. 

\subsection{Broken particle-hole symmetry}

In real materials
topological insulator (TI) surface states are not simply described by massless Dirac cones. 
ARPES experiments show surface states with significant band-bending
and broken particle-hole
symmetry with respect to the band crossing point \cite{hasan, ren, Bi2Te3}. 
For this reason, Hamiltonian \eq{H} is the relevant low energy Hamiltonian to describe
topological insulator surface states. 
This has been confirmed by symmetry arguments \cite{Bi1, Bi2}. In \tab{arpes}, we show some of the values for the mass $m$, and the Fermi velocity $v_F$ characterising the Hamiltonian \eq{H} measured by ARPES. 
It is clear from this data that for typical doping levels, the topological insulator surface states are truly mixed normal-Dirac systems, and cannot be considered to be approximating either regime. 

\begin{table}[h]
\centering
\begin{tabular}{c|  c | c | c | c | c}
  \hline  \hline                      
  Material & $m/m_e$ & $v_F$ (ms$^{-1}$) & $\mu$ (meV) & $\eta$ & Ref. \\ \hline
  Bi$_2$Se$_3$ & 0.25 & $5.0\times 10^5$ & 300 & 0.54 &  \onlinecite{hasan}\\
  Bi$_2$Te$_2$Se & 0.13 & $3.4\times 10^5$ & 130 & 0.39 & \onlinecite{ren}\\
  Bi$_2$Te$_3$ & 3.8 & $3.9\times 10^5$ & 250 & 0.93 & \onlinecite{Bi2Te3}\\
  Bi$_{1.5}$Sb$_{0.5}$Te$_{1.7}$Se$_{1.3}$ & 0.32 & $4.6\times 10^5$ & 140 & 0.73 & \onlinecite{ando3}\\
  \hline  \hline
\end{tabular}
\caption{Estimates of model Hamiltonian
parameters for several 3D topological insulator materials.
Normal fermion mass $m$ and Dirac fermion Fermi velocities $v_F$
are estimated from ARPES. At the values of
the chemical potential $\mu$ quoted in these works, we have also calculated the parameter $\eta$, defined in equation \eq{eta}. 
}
\label{arpes}
\end{table}

The effect of broken particle hole symmetry on the transport properties of topological insulators has been investigated theoretically,\cite{shafftheory} and experimentally. \cite{shaffexp}

We can think of the two limiting cases of \eq{etalims} in terms of particle-hole symmetry. In particular, pure Dirac fermions have particle-hole symmetry for all $\mu$, whereas pure normal fermions have two degenerate bands and so have broken particle-hole symmetry for all $\mu$. For $0<\eta<1$, the system is neither purely Dirac nor purely normal, but is a mixture of the two. In particular, near $k=0$, for any finite $\eta$, the electrons are Dirac fermion-like, and far away from $k=0$, they become normal fermion-like. We can rearrange \eq{etalims} to quantify this statement in terms of $\mu$

\beq
\frac{\mu}{mv_F^2}=\frac{1}{\eta} -1.
\eeq
We can say that $\eta$ provides a measure of broken particle-hole symmetry with respect to $k=0$. The effect of broken particle-hole symmetry on the band-structures is elucidated in \fig{phfig}.

\begin{figure}[tbp]
\centering\includegraphics[width=8.6cm]{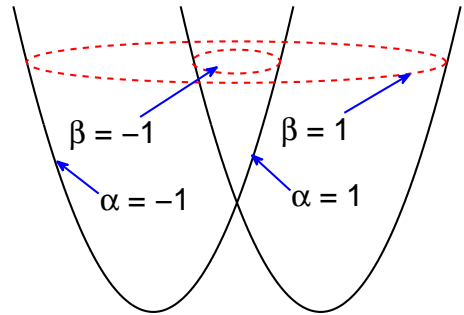}
\caption{For all energies, there are two distinct cyclotron orbits (Fermi surfaces). However, for real materials the Hamiltonian \eq{H} is an effective one for low energies, and the outer orbit may not actually exist. Nevertheless, for systems with small $\eta$, such as Rashba systems,\cite{rmpspintronics} there are two possible orbits, indexed by $\beta = \pm1$ for outer and inner orbits. In the figure, $\mu>0$, so the inner orbit ($\beta = -1$) circulates around the $\alpha = 1$ band, and the outer ($\beta = 1$) orbit circulates around the $\alpha = -1$ band. For $\mu<0$, both orbits circulate around the $\alpha = -1$ band. 
}\label{innerouter}
\end{figure}

\subsection{Gap opening: Zeeman and tunnel splitting, and beyond}

Mikitik and Sharlai have argued that, in general, for Bismuth-based 3D topological insulators, $\gamma = 0$ in the weak field limit.\cite{MS2012} It has been shown that the symmetries of the Bismuth based TIs don't allow for a non-zero gap $\Delta$ in \eq{H} \cite{Bi1, Bi2}, and so apart from the Zeeman splitting, and as we shall show, $\gamma$ is strictly zero.

However, it is clear from Table \ref{gammatable} that $\gamma$ is rarely observed to be zero, but in fact deviates substantially from zero. This implies that the g-factor may be $\sim 10-100$ in topological insulator surface states \cite{natphys, ando}. This corresponds to a Zeeman gap of $6-60$ meV in a $10$T field, which, when compared to typical Fermi energies of $50-300$ meV (Table \ref{gammatable}), is not insignificant. 

In topological insulator thin films, the surface states on the two faces of the sample hybridise, and thus open a gap \cite{gaptheory}. So even in Bismuth-based TIs, a non-zero $\Delta$ in \eq{H} can occur. In fact, thin films below the critical thickness (6 quintuple layers \cite{critical6}) have already been successfully produced in the lab, and the gap observed \cite{gapexpt1, gapexpt2, ando3}.

There are other mechanisms for gap formation which have been suggested, and even realised, including excitonic pairing gaps \cite{moore1, moore2} and a possible Higgs-type mechanism near the phase transition from the topological insulating phase to the topologically trivial insulating phase.\cite{acquisition}
 The formation of a gap is also crucial to various topological effects becoming manifest, including the as-yet unverified topological magneto-electric effect \cite{TME}.

\section{Quantum oscillations in topological insulator surface states}

\begin{figure}[tbp]
\centering\includegraphics[width=8.6cm]{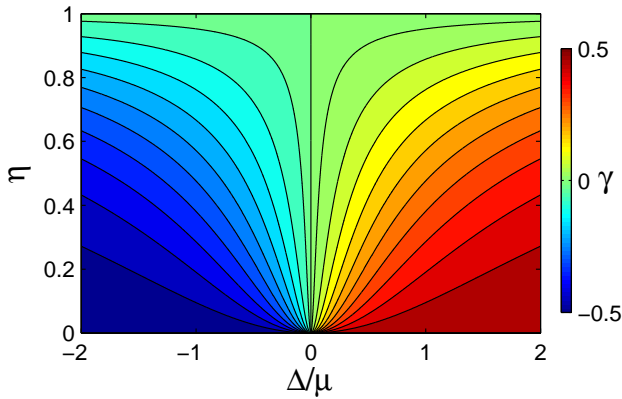}
\caption{(Color online) The phase offset 
$\gamma_+$ as a function of $\Delta$, the gap parameter, and $\eta$, the dimensionless parameter that tunes between normal fermions ($\eta = 0$) and Dirac fermions ($\eta = 1$). It is this quantity, and not simply the Berry's phase, that determines the phase offset in quantum oscillation experiments. If \emph{either} particle-hole symmetry is maintained ($\eta = 0,1$), or the system is gapless, the phase offset is a constant. If neither condition applies, the phase offset is a continuous function of $\eta$ and $\Delta$. The solid lines denote constant values of $\gamma_+$. }
\label{gamfig}
\end{figure}

\subsection{Non-universal phase offset in quantum oscillations}

The Berry's phase is given by \eq{berry}, and is readily calculated for the Hamiltonian \eq{H}. In particular, noting that the Berry's phase is a function of the wavefunctions only, the $k^2$ term in \eq{H} does not contribute for a path at fixed momentum space radius $k$, and so at all $\eta$, we have 

\beq
\Gamma_\alpha(k) = \pi \alpha(1 - \frac{\Delta}{\sqrt{\Delta^2 + \hbar^2 v_F^2 k^2}}),
\eeq
which is the same result obtained in Ref.[\onlinecite{Mx}] for the case $m\rightarrow \infty$. At fixed energy however, $k(\epsilon)$ is determined from the full dispersion \eq{dispn}, and the Berry's phase at fixed energy is implicitly dependent upon the full Hamiltonian. Moreover, the $\alpha = -1$ band has both an inner and an outer orbit, as shown in \fig{innerouter}. In particular, solving the dispersion at fixed energy $\epsilon$ for $k^2$, we obtain

\beq
k^2_\beta = \frac{2m}{\hbar^2}(\epsilon + mv_F^2 + \beta\sqrt{\Delta^2 + 2\epsilon m v_F^2 + m^2v_F^4}),
\eeq
where $\beta = \pm1$. The Berry's phase and phase offset $\gamma$ will then acquire an inner/outer orbit index. For $\alpha = 1$, $\beta = -1$ always, as there is no outer orbit in this case. However for $\alpha = -1$, and $\epsilon <0$, $\beta = [-1,1]$ coexist, and for $\epsilon>0$, $\beta = 1$ always. At fixed energy then, 

\beq
\Gamma_{\alpha,\beta}(\epsilon) = \pi\alpha\Biggl[1 - \frac{\Delta}{mv_F^2(1 + \beta\sqrt{1 + \frac{2\epsilon}{mv_F^2} + \frac{\Delta^2}{(mv_F^2)^2}})}\Biggr]
\eeq
With the Berry's phase known, we are able to find the phase offset for quantum oscillations from \eq{gam}. For the inner orbits ($\beta = -1$), the relevant case for topological insulators, we obtain

\beq
\gamma_{\alpha,-}(\epsilon) =   \frac{\alpha\Delta}{2mv_F^2\sqrt{1 + \frac{2\epsilon}{mv_F^2} + \frac{\Delta^2}{(mv_F^2)^2}}}.
\label{gam2}
\eeq
This is the central result of this paper, which shows that the phase offset expected in quantum oscillation experiments on topological insulator surface states is non-universal, but in fact depends upon the energy scales of the gap, and a mixed normal-Dirac fermion energy scale, $mv_F^2$. 

For the outer orbits, only relevant for the $\alpha = -1$ band, and, for example, Rashba systems where $v_F$ is small,\cite{rmpspintronics} 

\beq
\gamma_{-,+}(\epsilon) =  \gamma_{-,-}(\epsilon) + \frac{\Delta mv_F^2}{(x + \frac{mv_F^2}{2})^2 - \frac{m^2 v_F^4}{4}}.
\label{gam2outer}
\eeq
Where $x = \sqrt{\Delta^2 + 2\epsilon m v_F^2 + m^2 v_F^4}$. We have included this result for completeness, but from now on will focus on the inner orbits, relevant for topological insulator surface states, and drop the $\beta$ index in \eq{gam2}.

There are three particularly interesting limiting cases:

\begin{equation}
\gamma_\alpha(\epsilon)= \left\{
\begin{array}{rl}
\frac{1}{2}, & \quad v_F\rightarrow 0 \,\,\,\mathrm{(Normal}\,\,\mathrm{fermions)} \\
0, & \quad m\rightarrow \infty \,\,\,\mathrm{(Dirac}\,\,\mathrm{fermions)}\\
0, & \quad \Delta\rightarrow 0 \,\,\,(\mathrm{Mixed}\,\,\mathrm{gapless})
\end{array} \right.
\end{equation}
The second and third limit together are particularly interesting, as they imply that for infinitesimally small $v_F$, at $\Delta = 0$, $\gamma = 0$, and so $\gamma$ jumps discontinuously from $0$ to $1/2$ at $v_F = 0$. This discontinuity is apparent in the vertical line of \fig{gamfig} at $\Delta = 0$ as $\eta\rightarrow 0$.

The second limit gives the same result obtained previously by 
Fuchs \emph{et al.}.\cite{Mx}: in the presence of particle-hole symmetry
and a Dirac gap the phase offset is zero, even though the Berry's phase $\Gamma$ is not equal to $\pi$ and varies with the chemical potential.


\fig{gamfig} shows the expected phase offset for the electron band of gapped systems with both normal and Dirac components in terms of the two parameters $\tilde{\Delta}=\Delta/\mu$ and $\eta$. The phase offset \eq{gam2} can be written in terms of these as follows:

\beq
\gamma_\alpha(\Delta/\mu,\eta) = \frac{\alpha (1 - \eta)\Delta/\mu}{2\sqrt{\eta(2 - \eta) + (\Delta/\mu)^2(1 - \eta)^2}}.
\eeq
The phase offset varies as a function of the energy scale $mv_F^2$. The horizontal line $\eta = 0$ corresponds to purely normal fermions, and $\eta = 1$ to purely Dirac fermions. We can see that along these two lines, $\gamma$ is indeed quantized to $1/2$ and $0$ respectively, as expected. The topological result of Fuchs \emph{et al.} \cite{Mx} can be seen along the horizontal $\eta = 1$ line of the upper panel. Clearly in the more general case, the same topological arguments made in that case, relating $\gamma$ to a winding number, no longer apply, as $\gamma$ is no longer quantized. Moving away from the horizontal line at $\eta = 1$ corresponds to the breaking of particle-hole symmetry. Moving away from the line $\Delta = 0$ can correspond to breaking inversion or time reversal symmetry, depending on the basis, if the low energy theory \eq{H} is applicable at a high symmetry point in the Brillouin zone. In this particular regime, and considering just a single band, we recover the result of Mikitik and Sharlai,\cite{MS2012} that $\gamma = 0$ irrespective of the extent of particle-hole asymmetry. However, in more general systems, this does not correspond necessarily to the breaking of any particular symmetry. A prominent example is graphene with a staggered sub lattice potential, where the symmetry being broken is a sublattice one \cite{gappedgraphene}. 

So we see that in the general case, if either of the two symmetries are retained, such that $\Delta = 0$ or $\eta = 0$ or $1$, $\gamma$ is a constant, and independent of energy. If both are simultaneously broken, then the phase offset is a function of the gap (external field), the chemical potential, and the normal-fermion--Dirac fermion energy scale $mv_F^2$.

\subsection{Cyclotron effective mass}
Using the temperature dependence of the magnitude of the magnetisation at fixed field, one can fit the prefactor in \eq{M} that is $\lambda/\sinh(\lambda l)$, where $\lambda$ is given by \eq{lambda}. Usually, one can then determine the quantities $m$ (normal fermions) or $v_F$ (Dirac fermions). However, in our mixed system, the cyclotron
effective mass is more complicated, being given by \eq{mass}, where, reintroducing the inner-outer orbit index, 

\beq
S_{\alpha,\beta}(\epsilon) = \frac{2\pi m}{\hbar^2}\biggl[\epsilon_\alpha + mv_F^2\biggl(1 + \beta\sqrt{\frac{\Delta^2}{(mv_F^2)^2} + \frac{2\epsilon_\alpha}{mv_F^2} + 1}\biggr)\biggr],
\label{S2}
\eeq
as outlined in \fig{innerouter}.

We can easily check that as $v_F\rightarrow 0$, we obtain the normal fermion result of $S'(\epsilon) = 2\pi m$, and by expanding to all non-zero orders as $m\rightarrow \infty$, we obtain the usual Dirac fermion result of $S'(\epsilon) = \pi\epsilon/v_F^2$.

\begin{figure}[tbp]
\centering\includegraphics[width=8.6cm]{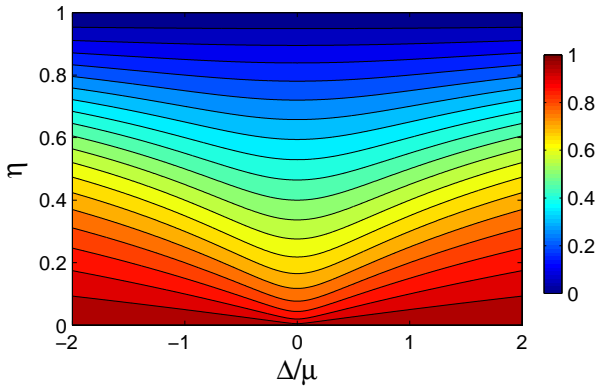}
\caption{(Color online) The relative cyclotron mass $m^*/m$ as a function of $\Delta/\mu$ and band bending. At $\eta = 0$, $m^* = m$. At $\eta = 1$, $m\rightarrow \infty$, so $m^*/m\rightarrow 0$. As the gap is decreased, or the chemical potential increased, the system becomes more Dirac-like. At $\eta = 0$ or $\eta = 1$, the cyclotron mass is independent of the gap/chemical potential, however in between the two this is not the case. }
\label{effectivemass}
\end{figure}

However, we note that in the mixed case, experimentally determining \eq{S2} does not directly give $m$ or $v_F$, but contains an expression with a pair of energy scales $mv_F^2$, and $\Delta$. In fact, for a mixed system such as this, we obtain from \eq{M}, $\lambda \propto \frac{dS}{d\epsilon} = 2\pi m^*$, and so we have

\beq
\frac{m^*}{m} = 1 - \frac{mv_F^2}{\sqrt{\Delta^2 + 2\epsilon mv_F^2 + (m v_F^2)^2}}.
\label{mst}
\eeq
In \fig{effectivemass} we have shown the ratio of the cyclotron mass to the normal fermion mass, the relative cyclotron mass, given by

\beq
\frac{m^*}{m} =1 - \frac{\eta}{\sqrt{\frac{\Delta^2}{\mu^2}(1 - \eta)^2 + \eta(2 - \eta)}} . 
\eeq
This figure shows the gap and chemical potential dependence of the cyclotron mass, highlighting that decreasing the gap or increasing the chemical potential tunes the cyclotron orbit to a more Dirac-like part of the spectrum. It also demonstrates that the interpolation between normal and Dirac fermions smoothly interpolates the relative cyclotron mass between 1 and 0. Note that the vanishing of the relative cyclotron mass at $\eta = 1$ is due to $m\rightarrow \infty$. In the Dirac limit, the cyclotron mass is $m^* = \epsilon/v_F^2$. Nevertheless, at either $\eta = 0$ or $\eta = 1$, the effective mass $m^*$ is gap and chemical potential independent. In between however, this is not the case, and the effective mass is a function of the gap and the chemical potential.

\subsection{Quantum oscillations and Landau level index plots}
The oscillations observed in resistivity (\eq{resistivity}) and magnetisation (\eq{M}) experiments have extrema at completely filled or empty Landau levels, to which we can assign integers and half integers $n\,\,(n+1/2)$. Therefore a plot of the location of the minima/maxima as a function of magnetic field can be used to identify the magnetic field values at which one has a filled/half-filled Landau level. From \eq{resistivity} and \eq{M}, it is clear a sensible ordinate for a plot of minima/maxima is the inverse field strength, $1/B$. To a good approximation then, (\emph{i.e.} only taking the first term in the series in \eq{M}), the condition for integer filled Landau levels is 

\beq
\frac{S_\alpha(\epsilon)}{2\pi \hbar e B} - \gamma_\alpha(\epsilon) = n - \Lambda,
\label{condition}
\eeq
where $\Lambda = 1/2 (0)$ for the minima (maxima) of the longitudinal resistivity, and $\Lambda = 3/4 (1/4)$ for the minima (maxima) of the magnetization. Further, we note that as $1/B\rightarrow 0$, the intercept of the index plot with the $n-$axis (offset by $\Lambda$) gives $\gamma(\epsilon)$. 

\begin{figure}[tbp]
\centering\includegraphics[width=8.6cm]{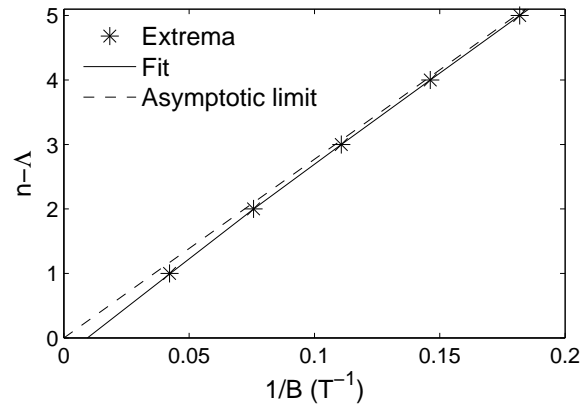}
\caption{(Color online) Robust determination of $\gamma_{B\rightarrow0}$ from an index plot. The crosses represent the extremum of the resistivity or magnetisation from \eq{resistivity} or \eq{M} corresponding to filled Landau levels. The solid line is the small field fit, \eq{largen}. It is clear that even for a system with pronounced Zeeman splitting (here $g_s = 50$), the fit is reliable at all filling factors. The dashed line is the linearized, asymptotic fit ($d\gamma/dB=0$ in \eq{largen}), which is asymptotically exact as $n\rightarrow \infty$. Extrapolating this line back to $1/B\rightarrow0$ gives the topologically relevant phase offset, $\gamma_{B\rightarrow0}$, which shows the Dirac nature of the surface states. System parameters are relevant to Bi$_2$Te$_2$Se \cite{ren}: $v_F = 3.4\times10^5ms^{-1}$, $m = 0.13m_e$, $g_s = 50$. }
\label{fanfig}
\end{figure}

In order to determine the index plot, a knowledge of both the area of the orbit, $S(\epsilon)$, and the phase offset $\gamma(\epsilon)$ is required. We have already 
given $\gamma(\epsilon)$ in \eq{gam2}, and plotted $\gamma$ for all $\eta$ and a range of gap parameters in \fig{gamfig}. $S(\epsilon)$ is 
given
in \eq{S2}, however it is interesting to note the limiting cases:

\begin{equation}
S_\alpha(\epsilon)= \left\{
\begin{array}{rl}
2\pi m \epsilon/\hbar^2, & \,\, v_F\rightarrow 0 \,\,\,\mathrm{(Normal\,\,\,fermions)} \\
\pi(\epsilon^2 - \Delta^2)/\hbar^2v_F^2, & \,\, m\rightarrow \infty \,\,\,\mathrm{(Dirac\,\,\,fermions)}\\
\end{array} \right.
\end{equation}
Firstly, we note that $\gamma$, \eq{gam2}, only varies as a function of the gap parameter in mixed normal-Dirac systems, being constant with respect to the gap in purely normal or purely Dirac systems. Therefore, for purely normal or purely Dirac systems, one expects the extrapolation of an index plot as $1/B\rightarrow 0$ to give $\gamma(\epsilon) = 0$ or $1/2$. Second, we notice that for gapless Dirac fermion and all normal fermion systems, $S(\epsilon)$ does not vary with the external field, and so the index plot is strictly linear. For a Zeeman gapped massive Dirac fermion system however, there is a nonlinearity in the index plot due to the Zeeman term $\Delta = g_s \mu_B B$. Fortunately for a material such as graphene, there is no Zeeman gap, and so the index plot reliably gives the expected intercept for relativistic fermions.\cite{novogeim} In all these cases then, correctly extrapolating the intercept at $1/B\rightarrow 0$ yields a constant field and energy independent value of $\gamma$. 

In a system with non-zero Zeeman splitting, and both normal fermion and Dirac fermion terms however, $\gamma(\epsilon)$ becomes field and energy dependent, as does $S(\epsilon)$. When calculating a Landau level index plot from the maxima/minima of the oscillations (\eq{M}) in mixed systems such as topological insulator surface states then, the index plot as a function of $|B|$ becomes uniquely nonlinear according to the energy scales of the Zeeman splitting and, as is clear from \eq{S2} and \eq{gam2}, $mv_F^2$. In \fig{fanfig} this nonlinearity with increasing field can be clearly observed, and has been pointed out elsewhere \cite{ando,SWP}. 

It is clear from the above discussion that extrapolating an index plot to $1/B\rightarrow 0$ yields an estimate of $\gamma(\epsilon)$. However, from \eq{gam2}, we see that $\gamma(\epsilon)$ in the case of mixed normal fermion-Dirac fermion systems depends on the energy scales $\Delta$ and $mv_F^2$. For this reason, the index plot cannot possibly discriminate between $m$ and $v_F$. The cyclotron
effective mass (\eq{mass}), on the other hand, intrinsically bears such a distinction. Through a combined approach of measuring the temperature dependence of the 
oscillation amplitudes, and thus obtaining $m^*$ (\eq{mass}), together with index plot measurements to determine the energy scales $mv_F^2$ and $\Delta$, one can determine both the normal fermion mass $m$, and the Dirac fermion velocity $v_F$. 


\section{Using Landau level index plots to extract the phase offset}
\subsection{Taking into account the effect of a bulk Fermi surface}
The quantum oscillation experiments performed thus far \cite{qo1, qo2, natphys, ren, ando3, mol, veldhorst, nanoL,xiong, xiong2, natcom,ando33}, tend to report a substantial bulk contribution to the measured resistivity. Although there is a bandgap at the $\Gamma$ point of the surface Brillouin zone, as clearly observed with ARPES\cite{hussain,hasan,hasan0}, the bulk 3D Brilluoin zone is not usually gapped throughout, but has a Fermi surface.

As our Hamiltonian, \eq{H}, is merely the effective surface theory of a bulk system, we must take into account the effects of the bulk on quantum oscillation experiments. In particular, we must consider the following expression:

\beq
N_e = \int_{-\infty}^\mu \rho_b(\epsilon)d\epsilon + \int_{-\infty}^\mu\rho_s(\epsilon)d\epsilon
\label{N}
\eeq
where $N_e$ is the number of electrons, which is a constant, $\rho_b$ is the density of states of the bulk bands, and $\rho_s$ is the density of states of the surface bands. As we switch on a magnetic field, $N_e$ remains constant, $\rho_s(\epsilon)$ thus changes due to the opening of a Dirac gap, and thus $\mu$ varies. Since the surface band is a two-dimensional band, of which there is only one, whereas the bulk bands are three-dimensional and there are many, the number of bulk carriers is much greater than the number of surface carriers. 

Therefore, there are two distinct regimes. 

\begin{enumerate}
\item \emph{If there is a bulk Fermi surface}, the contribution to $N_e$ will come almost entirely from the first term in \eq{N}, and so the chemical potential of the surface band will be dictated by the change in chemical potential of the bulk band. In the limit where the bulk Fermi surface shifts by a negligible amount, this is equivalent to taking the surface theory, \eq{H}, in the grand canonical ensemble, where $\mu$ remains constant, and the Fermi wavevector $k_F$, and therefore $S(\mu)$ varies. 
\item \emph{If the bulk is completely gapped}, the first, bulk, contribution to \eq{N} remains constant, and the change in $\mu$ is dictated entirely by the surface theory. This corresponds to the case most frequently considered, where it can be shown that the Fermi wavevector remains constant (Luttinger's theorem \cite{lutt}), and thus $S(\mu)$ remains constant as $\mu$ varies.
\end{enumerate}

Xiong \emph{et. al.}\cite{xiong} have pointed out a second subtlety in the presence of a bulk Fermi surface. They argue that when measuring the longitudinal resistivity, the contribution of the bulk resistivity demands that filled Landau levels be associated with resistivity maxima, rather than resistivity minima.

\subsection{Extracting the phase offset at small fields}
Experimental determination of the index plot in this case of mixed Dirac and normal fermions is, as we have discussed, fraught. In the presence of an unknown gap, an intercept with $n$ yields a non-quantized value of $\gamma$ which can sit anywhere on an equal-$\gamma$ contour in \fig{gamfig}. Furthermore, if the gap is from Zeeman splitting, it is linearly dependent upon $B$, and so, as pointed out by Taskin and Ando \cite{ando} as well as Seradjeh, Wu and Phillips \cite{SWP}, and shown in \fig{fanfig}, will not produce a straight line in the index plot as a function of inverse field strength. 

As the phase offset $\gamma(\epsilon)$ is only non-constant when the system is both gapped \emph{and} the electrons are mixed normal and Dirac fermions,  we can use this fact to determine whether the system has a gap at zero field. It is interesting to note that due to both of these stipulations being simultaneously required, this ability of quantum oscillations to extract an intrinsic gap is only possible in the unique mixed normal-Dirac systems we are considering here. 

In the case of a system with no intrinsic band-gap, we would like to measure $\gamma_{B\rightarrow 0} = 0$, the `topological' result for Dirac fermions. If the system has a tunnel-split gap, or some other more exotic gap, then $\gamma_{B= 0} \ne 0$.

In order to circumvent the difficulties outlined above, and in order to systematically obtain a reliable value for the topological phase offset, or a measure of the intrinsic band-gap, we propose the following procedure. Starting with \eq{condition}, we make a small field expansion and rearrange to obtain

\beq
n - \Lambda \approx \frac{B_0}{B} - \gamma_{B \rightarrow 0} - C\frac{d\gamma}{dB}\biggr|_{B\rightarrow0}B,
\label{largen}
\eeq
where 

\beq
B_0 = \frac{S(\epsilon)|_{B\rightarrow 0}}{2\pi \hbar e}
\eeq
is the area enclosed by the cyclotron orbit with no Zeeman correction, and
\beq
\frac{d\gamma_\alpha(\epsilon)}{dB}\biggr|_{B\rightarrow0} = \frac{\alpha g_s\mu_B}{2mv_F^2\sqrt{\frac{2\epsilon}{mv_F^2} + 1}}.
\label{dgam}
\eeq
The constant $C$ is, in general, unknown. In case 1 outlined in the previous subsection, where there is a bulk Fermi surface \emph{and} the change in $\mu$ with increasing magnetic field is negligibly small, $S(\epsilon)$ varies as a function of magnetic field. The variation of $S(\epsilon)$ with magnetic field can be approximated by expanding \eq{S2} with respect to the field. The second term in such an expansion, which goes as $B^2$, is $\propto d\gamma/dB|_{B\rightarrow0}$. In this case then, $C$ has a contribution from the gradient of $\gamma$, as well as a contribution from the change in the area of the Fermi surface at $\mu$, which is proportional to the gradient of $\gamma$ and the ratio of the Zeeman energy ($\Delta_Z = g_s\mu_BB$) to the cyclotron energy of the normal fermion ($\hbar\omega_0 = \hbar eB/m$). In case 2, $S(\epsilon)$ is a constant, and so $C = 1$. Therefore 

\beq
C= \left\{
\begin{array}{rl}
1 + \frac{\Delta_Z}{\hbar\omega_0}, & \,\,\, \mathrm{\mu\,\,constant\,\,(bulk\,\,Fermi\,\,surface),} \\
1, & \,\, S(\epsilon)\mathrm{\,\,constant\,\,(bulk\,\,gap).}\\
\end{array} \right.
\label{C}
\eeq
The topologically relevant quantity is $\gamma_{B\rightarrow0}$, which is the intercept of \eq{largen} with the $n-\Lambda$ axis in the limit $d\gamma/dB \rightarrow 0$.

Our proposal is thus as follows:
\begin{enumerate}
\item  Plot the experimentally obtained extrema in the resistivity or the magnetisation as a function of 1/B.  
\item Fit the data to the fitting function

\beq
n - \Lambda = \frac{B_0}{B} + A_1 + A_2 B,
\label{largen2}
\eeq
where $B_0$, $A_1$, and $A_2$ are constants. 
\item The asymptotic low field limit ($B\rightarrow0$) is equivalent to $A_2 \rightarrow 0$ and yields a straight line, whose intercept with the $n$ axis is $\gamma_{B\rightarrow0}$. Therefore, the topologically relevant phase offset is $A_1$. If $A_1 = 0$, the surface states are gapless, and contain a Dirac component with a Berry phase of $\pi$. If $A_1\ne 0$, then the system may have an intrinsic band-gap.
\end{enumerate}


\eq{largen}, together with theoretically obtained minima of \eq{resistivity} or \eq{M}, is plotted in \fig{fanfig} for typical parameter values. The fitting function is seen to be a good fit for all $n$. The topologically relevant asymptotic form of \eq{largen} has also been plotted in \fig{fanfig}, where  $\gamma_{B\rightarrow0} = 0$, as expected for topological insulator surface states with no intrinsic band-gap. 

We emphasise that producing a straight line fit to the experimental points, rather than following the procedure outlined above, will only yield a reasonable estimate of $\gamma_{B\rightarrow0}$ for very large $n$. As can be seen from \fig{fanfig}, the asymptotic approach of the data to the straight line fit is slow (approaching as $1/B^{-1}$). A more reliable method is to fit \eq{largen2} to the data, and in this way extract the topologically relevant zero field phase offset. 

\section{Comparison with existing experiments}


\begin{figure}[tbp]
\centering\includegraphics[width=8.6cm]{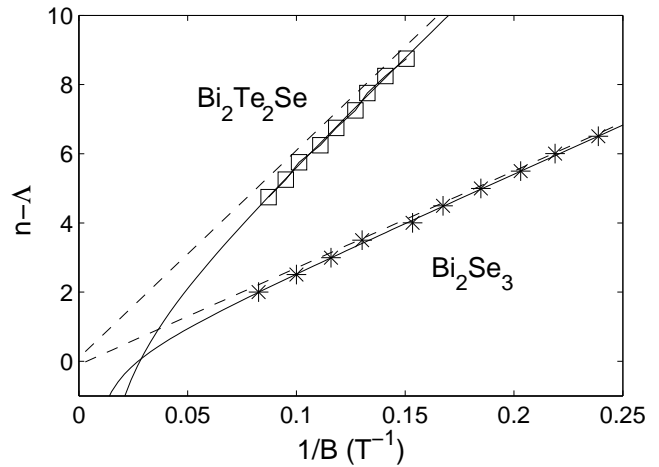}
\caption{
Determination of the zero field phase offset $\gamma_{B\rightarrow 0}$ using the fitting function \eq{largen2} (solid lines) on experimental data for Bi$_2$Te$_2$Se \cite{ren}, and Bi$_2$Se$_3$ \cite{qo1}. For Bi$_2$Te$_2$Se, we have shifted $n$ index by $1/2$, following Xiong \emph{et al.}\cite{xiong}, and obtain $\gamma_{B\rightarrow 0} = -0.1$, whereas a linear fit to the data yields $\gamma_{B\rightarrow 0} = -0.78$. For Bi$_2$Se$_3$, we obtain $\gamma_{B\rightarrow 0} = 0.1$, whereas a linear fit yields $\gamma_{B\rightarrow 0} = -0.36$. The dashed lines are the asymptotic plots of \eq{largen}, with $A_2 = 0$, whose $n-\Lambda$ intercept gives the topologically important zero field phase offset. 
}
\label{cffig}
\end{figure}

As has been pointed out by Taskin and Ando \cite{ando}, the nonlinearity of the index plots for large $B$ is evident in many existing studies, and the extrapolation to $1/B\rightarrow 0$ by fitting a straight line through the data points consistently yields $\gamma \ne 0,1/2$ (see \tab{gammatable}). It is also possible that for certain values of $g_s$, or external fields, the index plot will curve so much that the extrapolation spuriously yields $\gamma = 1/2$ for a topological insulator surface state. In fact Analytis \emph{et al.}\cite{natphys} obtained such a result with their sample 2 (see \tab{gammatable}). It is clear from our results that these findings are to be expected, and the problem may be systematically circumvented by fitting \eq{largen2} to the low field data and and thus obtaining a reliable estimate of $\gamma_{B\rightarrow0}$, as outlined above.

In \fig{cffig} we have performed just such an analysis for two existing experimental studies on  Bi$_2$Te$_2$Se \cite{ren}, and Bi$_2$Se$_3$ \cite{qo1}. Following the discussion of Xiong \emph{et al.}\cite{xiong}, we have shifted the index identification $n$ in the former by $1/2$. The reason for this being that the bulk contribution to the longitudinal resistivity causes the maxima, rather than the minima, to be associated with a completely filled Landau level. If this is the case, then a linear extrapolation of the data yields an intercept of $\gamma_{B\rightarrow0} = -0.78$. On the other hand, using \eq{largen2}, our fit becomes

\beq
n - \frac{1}{2} \approx \frac{60}{B} + 0.1 - 0.05 B
\eeq
Comparing with \eq{largen} then, we find $B_0 = 60$ T, in agreement with the value obtained in the original paper \cite{ren}, and $\gamma_{B\rightarrow 0} = 0.1$, which is close to the expected value of $0$. From \eq{dgam}, and the expression for $C$ at fixed $\mu$, \eq{C}, we see that if we use known values of $m$ and $v_F$, we can estimate the g-factor from the third term in the above result. Using this value we obtain $g_s\approx 60$ for $\eta = 0.39$, which is in broad agreement with similar estimates given elsewhere \cite{natphys,ando}.

Turning to the case of Bi$_2$Se$_3$ \cite{qo1}, we obtain

\beq
n - \frac{1}{2} \approx \frac{28}{B} - 0.1 - 0.018 B
\eeq
Again, we obtain a good agreement with $B_0 = 28$ T, and a zero phase offset of $\gamma_{B\rightarrow0} = 0.1$, whereas a linear fit yields $\gamma_{B\rightarrow0} = -0.36$. Using the material parameters of Bi$_2$Se$_3$ from \tab{arpes}, we obtain broad agreement with estimates given elsewhere, namely $g_s\approx 50$ for $\eta = 0.54$.
 
This brief comparison with experiment highlights the need for further experiments on higher purity samples with smaller Dingle temperatures. This will enable extending the index plots to lower fields.

\section{Application to spintronics}
For $\eta$ small, the Hamiltonian \eq{H} corresponds to a typical spintronics system: a two dimensional electron gas, such as GaAs and InAs quantum wells with Rashba spin-orbit interaction.\cite{rmpspintronics} For $v_F$ small, the two bands become nearly degenerate, and both contribute to quantum oscillations, \emph{i.e.} the sum over $\alpha$ must be completed in \eq{M}.

A robust treatment of these systems is not within the scope of the current work, as inter-band effects contribute a significant third portion to the oscillations \cite{MIS1,MIS2}, on top of the oscillations of the two intra-bands, and so a third term is required in \eq{M}, as inter-band tunnelling becomes possible.

However, the results of the present work do allow one particularly striking prediction for spintronic systems. According to \eq{gam2}, and \fig{gamfig}, for \emph{any} finite Dirac velocity, or in this particular case, Rashba spin-orbit interaction, as $\Delta\rightarrow 0$, $\gamma \rightarrow 0$. For very small Rashba terms though, this transition from $\gamma\approx \pm1/2$ to $\gamma = 0$ becomes almost step-like. Therefore, for all but the tiniest of Zeeman fields, one might expect to measure $\gamma = 1/2$, even though the $B\rightarrow0$ limit indeed gives $\gamma = 0$. For low fields, the intra-band components dominate \cite{MIS3}, and so it is expected that in the limit $B\rightarrow 0$, the model outlined here without inter-band scattering will become relevant. 

Experimental observation of an index plot with intercept consistent with $\gamma = 0$ in a 2DEG with small Rashba splitting would be a powerful confirmation of the current work.

\section{Conclusions}
We have developed a semi-classical theory of quantum oscillations in general particle-hole symmetry broken systems. In particular, we have applied the formalism to mixed normal fermion -- Dirac fermion systems, which is particularly relevant to topological insulator surface states. 

By properly treating the pseudo-spin magnetic moment, we have shown that the phase offset observed in quantum oscillation experiments is, in general, not quantized. We found that the quantized result can be expected in gapless systems, or in particle-hole symmetric systems, in agreement with previous studies. However, we have shown that if particle-hole symmetry is broken, and there is a gap, then the phase offset is not quantized. 

We have developed a protocol which allows one to determine the material properties of topological insulator surface states, in particular the normal fermion mass, the Dirac fermi velocity, the g-factor, and the intrinsic gap, by using quantum oscillation experiments. In particular, the effective mass cannot be naively applied to give the Fermi velocity, but is corrected due to Zeeman splitting and the energy scale $mv_F^2$. Furthermore, the observed nonlinear index plots can be fit to a simple function at small fields. From this, one can read off the phase offset at zero field, and directly determine the intrinsic gap, or else regain the `topological' result of zero phase offset, as expected where there is no mixing of normal and Dirac fermions.

The unique interplay of the normal fermion mass and the Dirac Fermi velocity to create an  energy scale $mv_F^2$, together with the Zeeman or intrinsic gap, add rich subtleties to quantum oscillation experiments that allow this already powerful tool to probe material properties in new ways.

In future studies we will focus on topologically significant gap phenomena, such as the topological exciton condensate,\cite{moore2} and how quantum oscillation experiments can be used to identify and probe these topologically nontrivial regimes.

\acknowledgements

We thank J. Kokalj for critical readings of the manuscript and helpful discussions. We also thank O. Sushkov, A. Hamilton, and T. Li for constructive feedback on the results, and especially J.N. Fuchs for providing critical feedback of the preprint. Financial support was received from a
UQ Postdoctoral Fellowship (ARW) and an Australian Research Council Discovery Project (RHM).

\end{document}